\documentclass[%
 reprint,
nofootinbib,
 amsmath,amssymb,
 aps,
floatfix,
]{revtex4-2}

\usepackage{graphicx}% Include figure files
\usepackage{dcolumn}% Align table columns on decimal point
\usepackage{bm}
\usepackage[dvipsnames]{xcolor}% bold math
\usepackage[final]{hyperref} % adds hyper links inside the generated pdf file1
\usepackage{ulem}
\hypersetup{
    colorlinks=true,       % false: boxed links; true: colored links
    linkcolor=blue,        % color of internal links
    citecolor=blue,        % color of links to bibliography
    filecolor=magenta,     % color of file links
    urlcolor=blue
}

\begin{document}

\preprint{APS/123-QED}

\title{From Champagne to Confined Polymer:\\Natural and Artificial Bubble Nucleation}% Force 

\author{Carlos Arauz-Moreno}
\email{c.arauz_moreno@icloud.com}
%\altaffiliation[Also at ]{Saint-Gobain Research Paris, 39 Quai Lucien Lefranc,
%Aubervilliers 93360, France}%Lines break automatically or can be forced with \\

\affiliation{%
 Univ. Grenoble Alpes, CNRS, LIPhy, 38000 Grenoble, France
}%

\affiliation{Saint-Gobain Research Paris, 39 Quai Lucien Lefranc,
Aubervilliers 93360, France}%Lines break automatically or can be forced with \\

\author{Keyvan Piroird}
\affiliation{Saint-Gobain Research Paris, 39 Quai Lucien Lefranc,
Aubervilliers 93360, France}%
%\affiliation{
\author{Elise Lorenceau}
\affiliation{Univ. Grenoble Alpes, CNRS, LIPhy, 38000 Grenoble, France
}%

\date{\today}% It is always \today, today,

\begin{abstract}
 
In this study, we present an experimental work on bubble nucleation and growth using a model system comprised of viscoelastic polyvinyl butyral confined in a Hele-Shaw cell geometry that is decompressed at elevated temperatures. The appearance and growth of bubbles are connected to the temperature-induced shift in chemical equilibrium experienced simultaneously by two gases present in the bulk. The latter becomes simultaneously oversaturated with water vapor and slightly undersaturated in air. Our bubbles grow with various shapes and sizes depending on the initial morphology of the nucleus or the presence of neighboring bubbles. For large nuclei, bubbles grow anisotropically because of contact line pinning.  The likelihood of nucleation is related to the amount of water dissolved in the bulk and the imposed temperature. Counter-intuitively, the number of nuclei whence a bubble can grow is inversely correlated with said temperature. In an analogy with champagne, we show that nucleation can either be natural, at trapped fibers or dust particles, or artificial, at crenels we purposefully made in the glass surface. Our results indicate that the growth rate of bubbles can be impacted by the nucleation mechanism. 

\end{abstract}

\maketitle

%\tableofcontents

\section{\label{sec:intro}Introduction}

Bubbles are magnificent objects that, for better or worse, are at the center of many important applications. For example, they  are extremely valuable as ultrasound contrast agents in diagnostic imaging, show potential for drug delivery~\cite{kooiman2014acoustic, roovers2019role}, are used to probe the local rheology of soft matter materials~\cite{zimberlin2007cavitation}, and may even dazzle the senses as in  champagne tasting~\cite{liger2008recent, liger2005modeling}. They can, at the same time, be extremely detrimental for they are at the root of injuries in decompression sickness~\cite{yount1976bubble, solano2015gas}, the delamination of thin films~\cite{berkelaar2015water}, and can induce early failure\textemdash or themselves constitute failure\textemdash as seen in solar panels or architectural safety glass~\cite{ kempe2014evaluation, martin2020polymeric}. In these examples, bubble nucleation and growth occur in a variety of materials - binary liquids, biomaterials, gels, polymers. In such non-newtonian fluids, bubbles exhibit particular behaviors such as interaction between neighboring bubbles, which may deform toward one another, as if \textit{attracting} each other~\cite{haudin2016bubble}, or may develop irregular shapes when the local deformation rate exceeds the relaxation time of the polymer~\cite{kundu2009cavitation, tabuteau2009microscopic}. It is therefore important to understand the physical processes at play in materials more complex than a simple liquid if we are to maximize their positive potential while minimizing their harmful effects.

With this context in mind, we studied the nucleation and growth of bubbles in a model system constituted by viscoelastic polyvinyl butyral (PVB) which we confined in between two glass slides in a geometry reminiscent of a Hele-Shaw cell or sandwich-like assembly. Experimentally, we triggered bubble nucleation and growth by decompressing the model system at elevated temperatures, thus generating oversaturation. The latter is mostly independent of the applied pressure because of confinement and is instead linked to the imposed temperature which shifts the chemical equilibrium of gases dissolved in the bulk of the PVB polymer. In this respect, there are two gases of interest in our system: \textcolor{blue}{\textit{water}} and \textcolor{red}{\textit{air}}, each one having their own and distinct thermodynamic preference\textemdash solubility wise\textemdash in the PVB polymer when heated. 

Post decompression, we observe that non-coalescing bubbles may form with a morphology that is linked to the size and/or shape of the nuclei or the close presence of neighboring bubbles. Bubbles that stem from small nuclei grow with a circular shape provided they are well-separated from other bubbles. Bubbles that originate from large nuclei grow with a certain degree of anisotropy that is reminiscent of the original contour of the nuclei because the initial contact line of the bubble remains pinned. Regardless of nuclei size/shape, bubbles may develop anisotropy when growing near other bubbles. Meanwhile, the likelihood of nucleation/growth is linked to the amount of water initially dissolved in the bulk or the imposed temperature. With respect to the latter, higher temperatures \textemdash wherein the polymer becomes softer\textemdash result paradoxically in seemingly fewer nuclei whence bubbles may grow. 

Overall, our results show that bubble nucleation can be subdivided into natural and artificial mechanisms in analogous fashion to champagne’s. In the natural case, bubbles nucleate in trapped matter at the glass/polymer interface such as speckles or fibers. In the artificial case, bubbles grow from imperfections we purposely made on the glass. While in both nucleation mechanisms bubbles exhibit the hallmark of diffusive growth, i.e., $R \simeq (kt)^{1/2}$\cite{epstein1950stability}, where $R$ is the bubble radius, $k$ is the growth rate coefficient, and $t$ the time, we found that $k$ is affected by the nucleation mechanism itself. In the natural case, $k$ is \textit{globally} the same for all bubbles, which reflects in a way the total oversaturation of the polymer. In the artificial case, however, the growth rate is strongly influenced \textit{locally} by the size of the crenel nucleus. 

Setting aside our experimental conditions, we believe our results hold general truths for bubble nucleation and growth in composite, layered assemblies wherein gases or foreign matter become easily trapped at interfaces or in substrate defects. Two prime examples include laminated safety glass and photovoltaic modules, both staples of modern life wherein the PVB polymer is sandwiched between two glass slides and where large variations in temperature are necessary to bond the polymer to the confining glass.

\section{\label{sec:mat}Materials \& Methods}

\subsection{PVB Polymer}

In this paper, we used RB41, an architectural grade of PVB produced by Eastman. This particular incarnation of the polymer has a well-documented chemistry~\cite{elziere2019supramolecular} and has been the subject of several studies covering mechanical response~\cite{hooper2012mechanical,stevels2020determination}, blast performance in glass~\cite{del2016determining}, rheological behaviour~\cite{arauz2022extended}, surface tension~\cite{morais2006evaluation}, and water sorption~\cite{arauz2023water}.

RB41 is a terpolymer of vinyl acetate (1-2\%wt), vinyl alcohol (18-20\%wt), and vinyl butyral (80\%wt) with triethylene glycol di(2-ethylhexanoate) as a plasticizer molecule (20-30\%wt). Because of the polar OH groups from the vinyl alcohol units, the overall polymer is highly hygroscopic. In terms of physical appearance, RB41 comes in the form of a solid, thin sheet having a nominal thickness of 760 $\mu$m. Lastly, as shown in fig.~\ref{fgr:exp_setup}A, the surface exhibits a random roughness of characteristic length (average depth) $e\sim 40 \mu$m that renders the polymer translucent.  

\subsection{Sample preparation} \label{subsec:samples}

The preparation of our samples included two steps: (1) properly conditioning the polymer bulk with gases and (2) confining the PVB polymer in between two glass slides in a Hele-Shaw cell geometry. 

\textbf{Conditioning.} The polymer is stored under a controlled atmosphere to fix the amount of dissolved gases in the bulk, in particular air and water vapor.  To this end, we stored the PVB polymer under humidity-regulated air at constant activity $\varphi=p_w/P_{Sat}(T)$, where $p_w$ is the vapour pressure in the atmosphere and $P_{Sat}$ is the saturation pressure at the temperature $T$. This was performed in either a climatic chamber or in a desiccant box with silica gel for at least 48hrs. Three activity levels were used leading to the designations of wet ($\varphi=0.5$), moist ($\varphi=0.25$), and dry PVB ($\varphi=0.05$) respectively. For simplicity, we treat air as nitrogen (N\textsubscript{2}, 78.1\% in air) and disregard other gases (e.g., Ar, CO\textsubscript{2})\footnote{We chose to work experimentally with atmospheric air because this gas is prevalent during the manufacturing of architectural glass and photovoltaic panels. While oxygen can be found in relevant quantities in the aforementioned gas, its presence is of minimal relevance in our experiments. First, atmospheric air contains 3.71 times more N\textsubscript{2} than O\textsubscript{2}\textemdash a dominating proportion when also considering that previous solubility studies have found that the PVB bulk absorbs only 1.5 times more oxygen than nitrogen~\cite{haraya1992permeation}. Second, like N\textsubscript{2}, O\textsubscript{2} is well above its critical point in our experiments, and thus, both gases can be treated as incondensable with an identical heat of solution sign~\cite{van1950influence, klopffer2001transport, rogers1985permeation}. When taken together, these two reasons imply that a bubble growing/dissolving in PVB with a mixture of N\textsubscript{2}/O\textsubscript{2} will largely be driven by N\textsubscript{2}. Finally, the remaining gases in atmospheric air are present in negligible amounts and are likewise absorbed in very small quantities by the PVB polymer.}. 

\begin{figure}
\includegraphics[width=0.49\textwidth]{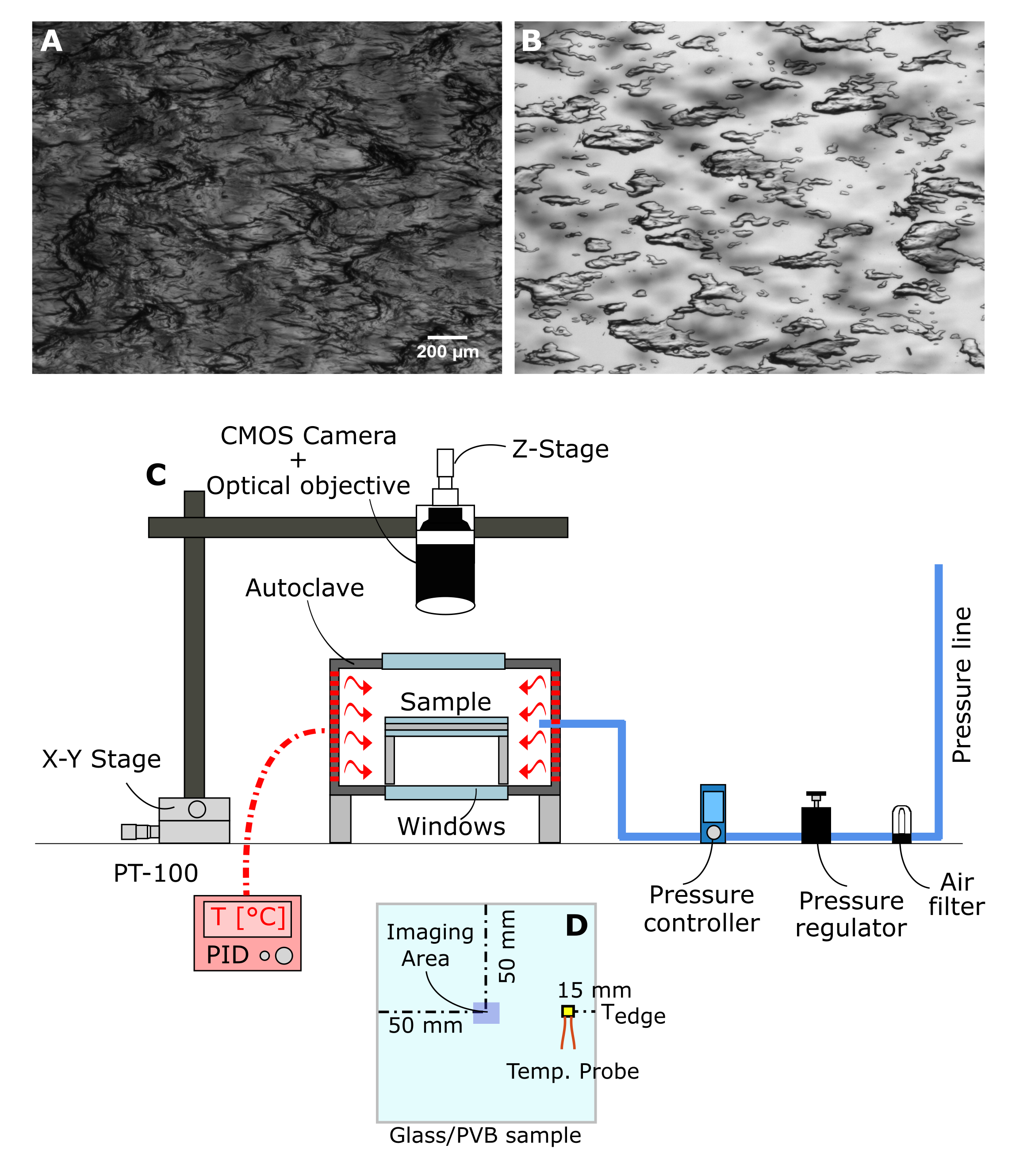}
\centering
\caption{Microscope images of polyvinyl butyral film when confined between two glass slides. \textbf{A} Original PVB surface roughness at room conditions. \textbf{B} Interfacial bubbles at the (top) glass/polymer interface after calendering. The shadowy regions are the unfocused bubbles from the bottom polymer/glass interface. \textbf{C} Experimental autoclave. \textbf{D} Sample schematics showing the imaging area (IA) whence the bubble measurements are taken.} 
\label{fgr:exp_setup}
\end{figure}

\textbf{Confinement.} We confined the polymer in a glass/polymer/glass sandwich (10 x 10 cm, with 2mm-thick glass). By slightly heating the sandwich (90°C, 15 mins) and subsequent calendering under light pressure (Linea DH-360, 1.4 m min\textsuperscript{-1}), the polymer roughness (fig.~\ref{fgr:exp_setup}A) partially melts away leaving behind anisotropic, interfacial bubbles (fig.~\ref{fgr:exp_setup}B). The polymer stiffness immobilizes theses interfacial bubbles at room conditions.

\subsection{Bubble experiments} \label{subsec:bubble_exps}

To study bubble nucleation and growth at high temperatures, we developed a transparent autoclave whose main body was constituted by a hollow cylinder made from stainless steel and a set of viewing windows (fig.~\ref{fgr:exp_setup}C). A pair of heating ropes (Omegalux, 400 watts/unit, not shown) were wound around the periphery of the cylinder to supply heat, while pressure was provided via a Fluigent pressure generator (2 bar) connected to a dedicated port in the autoclave. Images were taken at the center of the glass sample in a target imaging area (IA) using an overhead high-resolution camera (UI-3240ML) with a 75mm (or 105mm) optical objective (res.$\sim 10-18.6\mu $m px\textsuperscript{-1}), while light was shone from below the autoclave using an LED source (Schott KL 2500, not shown). During the experiments, the temperature was regulated using a PID controller and a thin film PT-100 sensor secured midway through the glass sample at 15 mm away from the right edge (fig.~\ref{fgr:exp_setup}D). 

The experimental protocol included the application of a heating ramp under a hydrostatic load, followed by rapid decompression of the autoclave chamber under isothermal conditions. The hydrostatic load was applied to dissolve the interfacial bubbles initially present in the sample (see fig.~\ref{fgr:exp_setup}B) inside the PVB polymer bulk. Meanwhile, the variation in temperature, from room conditions $T_i=25$°C to the temperature of decompression $T_{dec}$, triggered the nucleation and growth of bubbles post decompression by shifting the chemical equilibrium of the gases (\textcolor{red}{\textit{air}} and \textcolor{blue}{\textit{water}}) dissolved in the polymer bulk.  

\section{Results \& Discussion}

A bottle of champagne contains six times its volume in terms of CO\textsubscript{2}. This is one of the secrets behind its famous effervescence: the liquid is oversaturated~\cite{liger2008recent}. When the bottle is uncorked, the liquid \textit{releases} this excess gas to achieve chemical equilibrium with the surrounding air. Part of the CO\textsubscript{2} in the liquid escapes via the free interface and the remaining gas is available to form bubbles. Our situation is somewhat analogous. During conditioning, we set the initial amount of gas in the polymer bulk. By heating and compressing/decompressing, we shift the chemical equilibrium of the gases in the bulk, and thus mimic the effect induced by uncorking in champagne. In principle, bubbles should then grow (or not) depending on whether the PVB polymer \textit{releases} or \textit{absorbs} gases at the decompression temperature $T_{dec}$. 

For the growth of our bubbles, we considered the effect of different gas concentration levels in the polymer bulk (in particular water concentration), several temperatures (100,120,140°C), as well as the role played by nucleation itself. In doing so, we also gathered interesting information regarding their morphology. In this regard, our bubbles can be circular, elongated, or completely anisotropic as well.  

\textbf{Relative gas saturation}. Our post-decompression bubbles are inherently multi-component given the prevalence of air and water in the model system. Each gas has their respective relative saturation \textcolor{red}{$f_a$}, \textcolor{blue}{$f_w$} that is set independently by $f=1+\zeta$, where $\zeta = \Delta c/c_0$, and $c_0$ is the concentration at saturation conditions (room conditions in our case) and $\Delta c$ is the concentration difference relative to $c_0$ at the decompression temperature $T_{dec}$. Broadly speaking, $f>1$ means gas oversaturation and is linked to bubble growth, $f=1$ is the reference saturation state wherein gases are in equilibrium and bubbles are in principle static (whenever surface tension or other rheological effects are negligible), and $f<1$ implies gas undersaturation and concomitantly bubble shrinkage. 

Contrary to other decompression bubbles in the literature~\cite{gent1969nucleation, haudin2016bubble, penas2016history,kwak1983gas}, $f$ is in our case largely independent of the pressure drop (aside from possible entropic effects which we disregard) because our target area in the PVB polymer is closed off from the surrounding atmosphere given that (i) the confining glass is impermeable to our gases, and (ii) our samples are relatively large for the timescale set by diffusion. This also means that during our experiments the PVB polymer absorbs an excess quantity of gas only in accordance with that which was initially present in the interfacial bubbles themselves. The latter are comprised mostly of air because of the conditioning atmosphere that was used, i.e., $\textcolor{red}{x_a}=\textcolor{red}{n_a}/n\sim \textcolor{red}{p_a}/P_o\gtrsim0.99$, where $x_a$, $n_a$, $p_a$ are the mol fraction, number of moles, and partial air pressure in the conditioning atmosphere, and $P_o$ is the reference atmospheric pressure (101.325 kPa).  

We suppose the \textit{instantaneous} relative saturation $f$, before decompression, to be primarily influenced by how the Henry solubility constant ($H$, [$kg \ m^{-3}\ Pa^{-1}$]) of the gases evolves during our experiments. This is directly linked to their heat of solution $\Delta H_s$ in the PVB polymer and the shift in temperature from room conditions to $T_{dec}$. For water,  \textcolor{blue}{$\Delta H_{s,w}$}$ < 0$~\cite{arauz2023water}, whereas for air \textcolor{red}{$\Delta H_{s,a}$}$>0$. See the Appendix for a gravimetric study of our own where we show that $N_2$ (the proxy for air) behaves remarkably close as in  simpler, rubbery homopolymers such as polyisoprene in terms of sorption and diffusion.

For the relative saturation of water, we can disregard the excess amount of moisture from the interfacial bubbles seeing that this quantity is trivial compared to the ratio of the solubility constants. For example, with moist PVB ($\varphi=0.25$) in our experimental range (25-140°C), the relative water saturation is \textcolor{blue}{$f_w$}$\sim$\textcolor{blue}{$H_w$}$(T_i)/$\textcolor{blue}{$H_w$}$(T_{dec})\sim 1.1\times10^2$, i.e., the PVB polymer is in principle oversaturated with water vapour over 100 times! Evidently, the PVB polymer \textit{releases} water during our experiments, thus favoring the formation of bubbles. 

For the relative air saturation, we must first compute the total concentration of air in the system to then determine the effective solubility constant, and thus, the likelihood of air bubbles. Accordingly, we define the total air concentration $\textcolor{red}{c_a^*}$ [$kg \ m^{-3}$],

\begin{equation}
\textcolor{red}{c_a^*}=\textcolor{red}{c_d}+\textcolor{red}{c_i},
\label{eq:vol_air1}
\end{equation}

where $\textcolor{red}{c_d}$ is the concentration of air initially dissolved in the PVB bulk from natural saturation conditions and $\textcolor{red}{c_i}$ is the air concentration set by the initial volume of air present in the interfacial bubbles per unit volume of PVB. 

$\textcolor{red}{c_d}$ is given by

\begin{equation}
\textcolor{red}{c_d}=\textcolor{red}{H_a}(T_i)\textcolor{red}{p_{i,a}},  
\label{eq:vol_air2}
\end{equation}

with \textcolor{red}{$H_a$} being the solubility constant at the initial temperature $T_i$ while $\textcolor{red}{p_{i,a}}$ is the initial air partial pressure (conditioning pressure). 

Meanwhile, $\textcolor{red}{c_i}$ is determined as

\begin{equation}
\textcolor{red}{c_i}=A\frac{e}{h}\textcolor{red}{\rho_a}(T_i),  
\label{eq:vol_air3}
\end{equation}

wherein $A$ is the initial fraction area of interfacial bubbles, typically $\approx 0.32\pm0.02$SD in our experiments, $e$ is the characteristic polymer roughness, $h$ is the half-thickness of the PVB layer and \textcolor{red}{$\rho_a$}$(T_i)$ is the dry air density included for dimensional homogeneity with $\textcolor{red}{c_d}$.

Eq.(\ref{eq:vol_air1}) can conveniently be written as:

\begin{equation}
\textcolor{red}{c_a^*}=H_a^*\textcolor{red}{p_{i,a}},    
\end{equation}

with $H_a^*=$\textcolor{red}{$H_a$}($T_i$)$+A\frac{e}{h}\frac{\textcolor{red}{\rho_a}(T_i)}{\textcolor{red}{p_{i,a}}}$ being an effective solubility constant that accounts for the total amount of air in the system.

The relative air saturation at 100-140°C is then \textcolor{red}{$f_a$}$\sim$\textcolor{black}{$H_a^*$}$/$\textcolor{red}{$H_a$}$(T_{dec})\sim 0.92-1.1$, where we took \textcolor{red}{$\Delta H_{s,a}$}$\sim 5\times10^3$ J mol\textsuperscript{-1} (see the Appendix). Despite the excess air dissolved in the bulk from the interfacial bubbles, the PVB polymer is at most at saturation conditions since \textcolor{red}{$f_a$}$\sim1$. Therefore, pure air bubbles are thermodynamically unfavorable and should not form.

\begin{figure}
\includegraphics[width=0.5\textwidth]{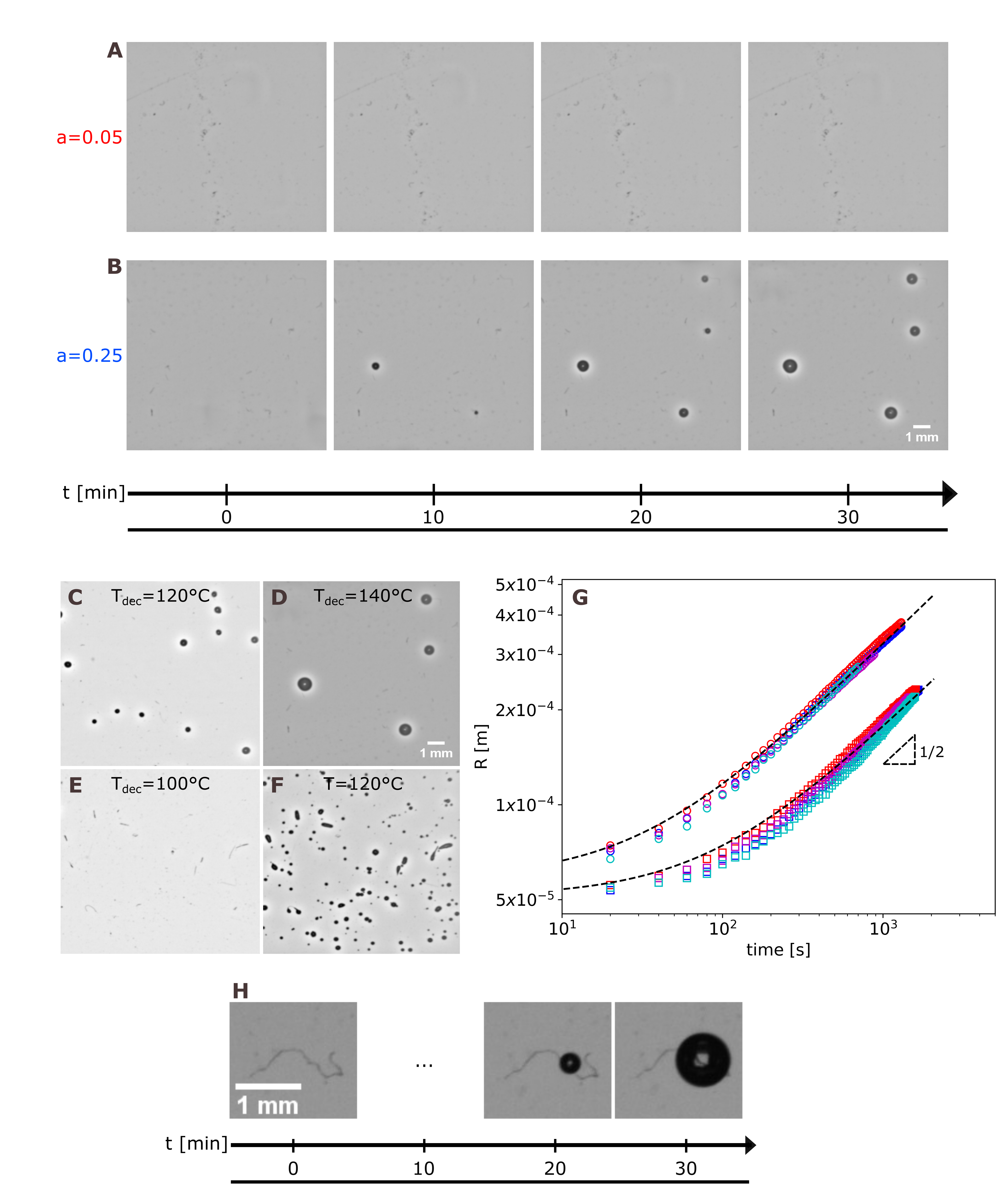}
\centering
\caption{Natural bubble nucleation and growth at elevated temperatures post decompression ($\Delta P=1$bar). \textbf{A} Lack of bubble growth with \textit{dry} PVB (\textcolor{red}{$\varphi=0.05$}). \textbf{B} Bubble growth with \textit{moist} PVB (\textcolor{blue}{$\varphi=0.25$}). \textbf{C}-\textbf{E} Image snippets of bubble growth at $T_{dec}=$120,140,100°C (note that figure D is the last panel from figure B). \textbf{F} Experiment from E at $T_{dec}$=100°C with subsequent increase in temperature to $T=$120°C. \textbf{G} $\square$\text{\large{$\circ$}}  Bubble radius as a function of time at 120, 140°C and associated diffusion model (eq.~\ref{eq:diff_model}) with the canonical signature $R\sim t^{1/2}$ (dashed lines). We restricted ourselves to bubble sizes $R\leq h$ where $h$ is the PVB half-thickness. The radii, as with other figures in the paper, was obtained automatically via imageJ for bubbles with an initial radius greater than or equal to 50$\mu$m. \textbf{H} Image sequence of natural bubble nucleation around a fiber ($T_{dec}$=140°C, $\Delta P=0.6$ bar).}
\label{fgr:reel_decompressionT}
\end{figure}

\textbf{Role of gas concentration}. Figs.~\ref{fgr:reel_decompressionT}A,B compare the bubbling behaviour between \textcolor{red}{\textit{dry}} and \textcolor{blue}{\textit{moist}} PVB. When using dry PVB, no bubbles nucleated after decompression in the time span of the experiment (fig.~\ref{fgr:reel_decompressionT}A). This meant that the air initially present in the anisotropic interfacial bubbles, and which was forcibly dissolved via the hydrostatic load, remained locked in solution inside the PVB bulk (\textcolor{red}{$\Delta H_{s,a}$}$ > 0$, \textcolor{red}{$f_a$}$\leq1$). Contrastingly, when using moist PVB, bubbles nucleated and grew (fig.~\ref{fgr:reel_decompressionT}B), which is consistent with water vapour escaping from the PVB bulk with temperature (\textcolor{blue}{$\Delta H_{s,w}$}$ < 0$, \textcolor{blue}{$f_w$}$>1$). Thus, post decompression, observing either bubble growth (or not) is linked to the initial amount of water in the polymer bulk.  

\textbf{Temperature effect}. In figs.~\ref{fgr:reel_decompressionT}\textbf{C}-\textbf{E}, we surveyed the effect of temperature on bubble nucleation and growth at 100, 120, and 140°C. As seen therein, the number of bubbles post decompression and their size was dependent on $T_{dec}$. In this regard, several bubbles, albeit smaller in size but in larger numbers, were observed at 120°C (fig.~\ref{fgr:reel_decompressionT}C) compared to 140°C (fig.~\ref{fgr:reel_decompressionT}D).

Puzzling, no bubbles nucleated in the run at 100°C (fig.~\ref{fgr:reel_decompressionT}E).

By continuing the latter experiment and increasing $T_{dec}$ from 100 to 120°C, we observed an explosive appearance of bubbles (fig.~\ref{fgr:reel_decompressionT}F). This suggests that the sample in fig.~\ref{fgr:reel_decompressionT}E had a large number of nuclei that only became active once we raised the temperature\footnote{We visualized our samples under a microscope to qualitatively compare the number of nuclei therein as a function of $T_{dec}$. To this end, we sampled the number of gas inclusions at the polymer/glass interfaces but disregarded the bubbles already detected by our autoclave system. Indeed, the number of said inclusions was inversely correlated with $T_{dec}$. However, ascertaining the quantitative relationship between the two was beyond the scope of the present work.}.  This leads to a twofold conclusion: (i) the number of nucleation sites post decompression is apparently negatively correlated with $T_{dec}$ (as set by the number of bubbles in figs.~\ref{fgr:reel_decompressionT}C,D,F) and (ii) seemingly inactive sites can be activated by increasing the temperature.  Most likely, the sites which we activated in fig.~\ref{fgr:reel_decompressionT}E by increasing the temperature to 120°C might have also become active at 100°C had we waited long enough. How long, however, is unknown.

We speculate there are two effects at play for the paradoxical relationship between $T_{dec}$ and the number of bubbles/nuclei. First, between 100-140°C, the rheology of the polymer varies significantly. For comparison, we estimate the shear modulus at 100°C ($t=1$s) to be $G\sim10^5$ Pa, which then decreases by an order of magnitude to $G\sim10^4$ Pa ($t=10^2$s). Contrastingly, at 140°C, the PVB polymer is between 3-15 times correspondingly softer~\cite{arauz2022extended}. We thus hypothesize that the high stiffness at 100°C probably restrained the nuclei from growing into visible bubbles, despite the relatively high water saturation ($\color{blue}{f_w}$ $\sim40$). Second, the higher the temperature, the more easily air can dissolve in the PVB bulk, thus diminishing the number of nuclei whence a readily visible bubble may eventually grow. More work is needed to conclusively verify the mechanism behind the number of nuclei that grow into visible bubbles.

\textbf{Diffusion model}. While bubbles nucleated at seemingly different times, their growth history collapsed onto a single curve, as portrayed simultaneously in fig.~\ref{fgr:reel_decompressionT}G for 120,140°C, when arbitrarily defining $t_o=0$ s for $R_o\geq 50 \mu$m. This selection of $t_o$ plays no role on the observed kinetics and any other bubble radius can be selected without affecting the results. We found that the growth curves of the bubbles at 120°C, 140°C are well described by a simple diffusion model~\cite{epstein1950stability} (dashed line in fig.~\ref{fgr:reel_decompressionT}G): 

\begin{equation}
R^2=kt+R_o^2,
\label{eq:diff_model}
\end{equation}

where $k$ is a growth rate constant that has units of a diffusion coefficient. Since at both 120, 140°C the bubble growth curves are self-similar, the growth coefficient $k$ is \textit{globally} the same for all bubbles. In turn, this means that our bubbles experience identical oversaturation conditions during their growth.  

The growth rate constant $k$ is impacted by temperature threefold. From the ideal gas law, the volume $V_B$ of the bubble is $V_B=n_B R_u T/P_B$, where $n_B, P_B$ are the number of moles and pressure in the bubble and $R_u$ is the universal gas constant. From this simple expression, we first conclude that the higher the temperature, the larger the bubble volume. At the same time, temperature affects the rate of $n_B$ which is a function of the diffusion $D$ and solubility $H$ coefficients that are themselves exponentially dependent on temperature. 

For the results provided in fig.~\ref{fgr:reel_decompressionT}G, the variation of the growth rate with temperature $k(140^{\circ}C)$/$k(120^{\circ}C)\sim 3.5$ is in the order of magnitude of the variation of $T_{dec}$ and the transport properties for water, which is respectively $\sim T_{140}/T_{120}\times D_w(140^{\circ}C)/D_w(120^{\circ}C)\times H_w(120^{\circ}C)/H_w(140^{\circ}C)\sim 3$ (see tables ~\ref{tbl:transport_comp}-\ref{tbl:model} in the Appendix for the transport constants). This confirms that our bubbles are globally driven by water vapour.

\textbf{Nucleation}. We hypothesized that small gas nuclei, whose size was below the resolution of our imaging system ($\sim10\mu$m), survived the application of hydrostatic pressure and then grew by mass diffusion after the pressure was released. These forms of nuclei are by no means exclusive to us – they have been observed in cavitation experiments for decades~\cite{liebermann1957air, atchley1989crevice, harvey1944bubble}. Such nuclei are, in fact, the second secret behind champagne's dazzling bubbles.

The level of CO\textsubscript{2} oversaturation in champagne (or of any other carbonated beverage for that matter) is not sufficient to form bubbles in the liquid bulk \textemdash the energy requirement is extremely high. Instead, during pouring, bubbles \textit{naturally} grow from nuclei trapped in cellulose fibers in the flute from the cleaning process, or \textit{artificially} from purposefully made imperfections on the glass, such as crenels or pits~\cite{liger2005modeling, liger2008recent}. The same nucleation mechanisms apply to us.

By verifying the experimental recordings, we observed that our post-decompression bubbles nucleated at what apparently were dust speckles or fibers trapped at the polymer/glass interface in agreement with natural nucleation. In fig.~\ref{fgr:reel_decompressionT}H a bubble is seen nucleating and growing from a fiber-like object in accordance with natural nucleation. In subsequent related experiments, we used a microscope to locate the bubbles and found them always at one of the polymer/glass interfaces but never in the PVB bulk, i.e., natural nucleation was exclusively of the heterogeneous kind, often with a clearly visible nucleation site at an interface. 

\begin{figure}[t]
\includegraphics[width=0.5\textwidth]{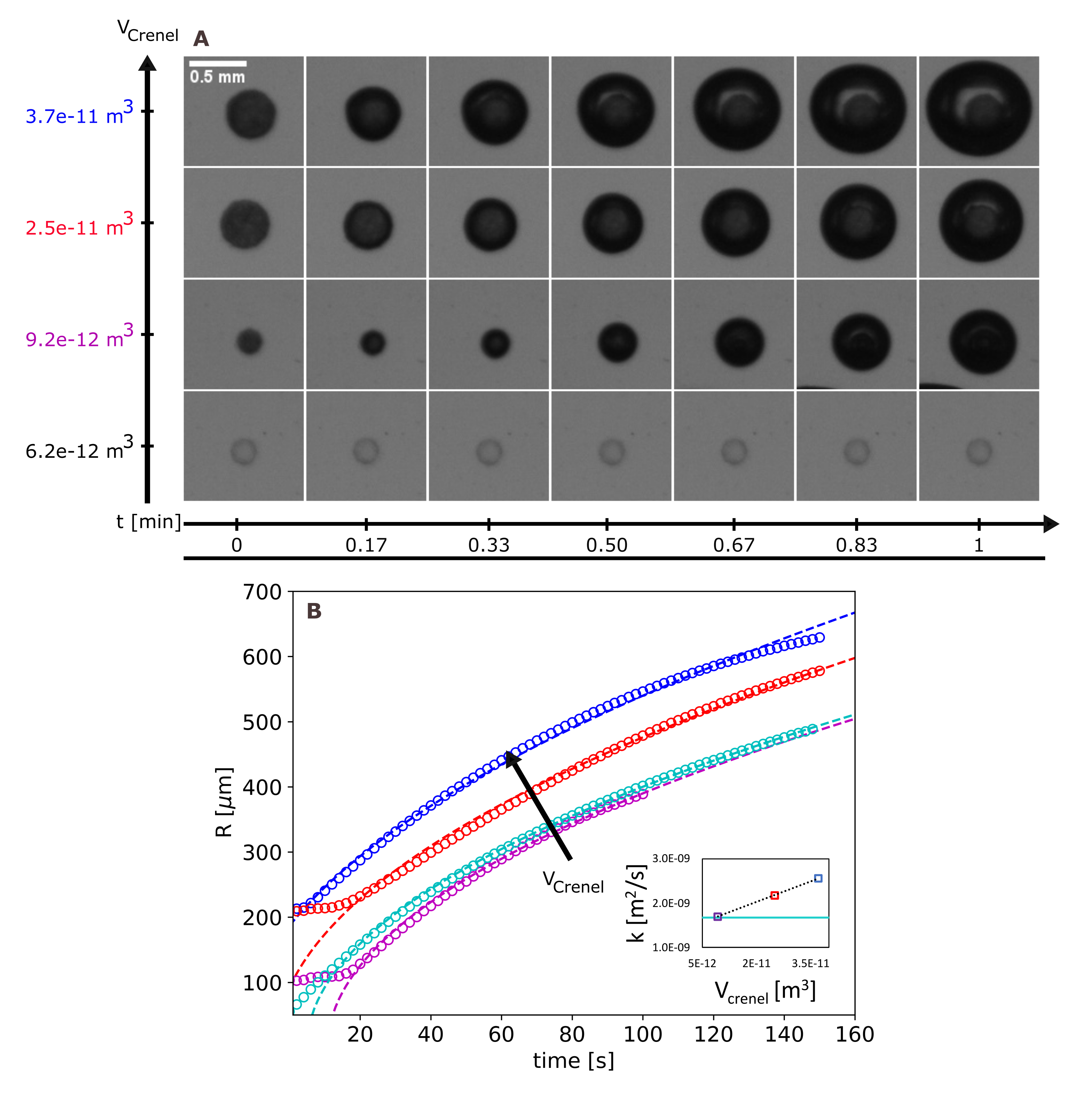}
\centering
\caption{\textbf{A} Artificial bubble nucleation and growth in moist PVB post decompression in four crenels of varying volume ($V_{crenel}$) etched on the glass surface ($T_{dec}\sim$144°C, $\Delta P=0.6$ bar). \textbf{B} \text{\large{$\circ$}} Blue, red, and purple circles indicate projected bubble radius $R$ versus time. The dashed line is a diffusion model $R\sim t^{1/2}$. The $R$ vs $t$ curves are color matched to the crenel volume in \textbf{A}. \textcolor{cyan}{\text{\large{$\circ$}}} Teal circles indicate bubble growth from natural nucleation. The inset plots the growth rate $k$ versus crenel volume. Symbols and solid line are color matched to their respective bubbles. The (teal) horizontal baseline is the growth rate obtained from a naturally nucleated bubble. The dashed black line is a linear regression.} 
\label{fgr:nucl_crenels}
\end{figure}

To induce artificial nucleation, we used a UV laser to pattern the surface of the glass with controlled defects which in our case took on the form of cylindrical crenels. In fig.~\ref{fgr:nucl_crenels}A, we present a subset of two sets of crenel having different radii $R$ (100, 200$\mu$m) and depths $h$ (196, 296 $\mu$m). These experiments highlight that bubble growth post decompression, beyond being globally dependent on water concentration in the PVB bulk and/or temperature, is also affected \textit{locally} by the amount of gas present in the glass imperfection itself ($V_{crenel}$).

In the smallest crenel, bubble nucleation and growth were not observed. In the remaining ones, however, bubbles grew immediately after decompression from their crenel nucleus. Their final size stood in direct relation to the initial gas volume present in the crenel itself (fig. ~\ref{fgr:nucl_crenels}B). The same seemingly applies to the growth rate $k$. As seen in the inset in fig.~\ref{fgr:nucl_crenels}B, $k$ starts from the baseline set by natural nucleation (solid teal curve) and thence evolves linearly with $V_{crenel}$. This marks a clear break from natural nucleation for there is now a local effect wherein each crenel bubble experiences different oversaturation conditions. 

The growth rate of the bubbles and power scaling, i.e., $R\sim t^{1/2}$, seems robust and bubbles maintained a circular shape, even when the bubble size surpassed the PVB thickness (760$\mu$m) in a short timescale. On the other hand, this means that these bubbles became pancake-like during our experiments.

\textbf{Bubble morphology}. In champagne, bubbles detach from their nucleation site and rapidly become spherical and ascend by buoyancy. The bubbles maintain sphericity because they are smaller than the capillary length and the fluid is newtonian. In non-newtonian fluids or brittle materials, by contrast, it has been shown that bubbles can take on complex shapes when the rate of deformation is larger than the polymer relaxation time~\cite{kundu2009cavitation, tabuteau2009microscopic}. Similarly, in a hydrogel, neighboring bubbles develop curved segments towards one another, whenever the distance separating them is roughly equivalent to their respective radii~\cite{haudin2016bubble}. In our case, the initial bubble morphology is linked to the shape of the nuclei before decompression or the presence of neighboring bubbles. 
\begin{figure}
\includegraphics[width=0.5\textwidth]{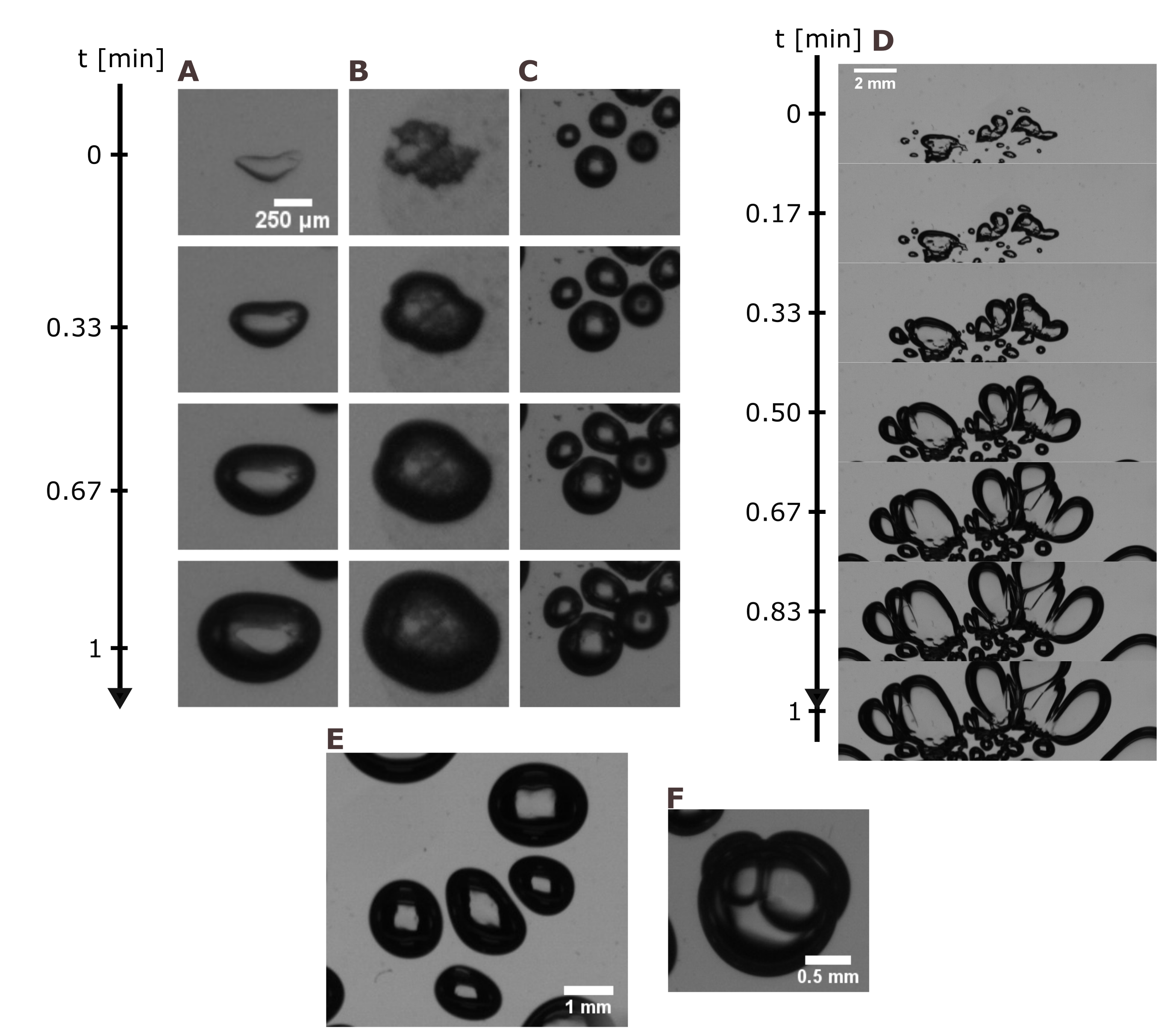}
\centering
\caption{Morphology of bubbles post decompression ($T_{dec}\sim140$°C,  $\Delta P=0.6$ bar). \textbf{A}-\textbf{B} Undissolved anisotropic bubbles (moist PVB) serve as nuclei for bubble growth. The latter grow with a contour resembling the original shape of the nucleus before approaching circularity as time passes. \textbf{C} Bubbles with an initial circular projection develop anisotropy when sufficiently close to neighboring bubbles (the sequence has been shifted by one minute to coincide with the scale from \textbf{A}-\textbf{B}). \textbf{D} A collection of partially dissolved anisotropic bubbles lead to irregular bubble morphologies (PVB conditioning $\varphi=0.5$). The anisotropy is not linked to the high water content (not shown) but rather to the presence of closely-packed anisotropic bubbles. \textbf{E}-\textbf{F} Examples of non-coalescing bubbles. In \textbf{E}, bubbles grew close to each other at the same polymer/glass interface and deformend as they came in close proximity. In F, three bubbles grew at different interfaces. The small bubbles grew at the bottom interface, while the large one is attached to the top interface.} 
\label{fgr:bubble_morphology}
\end{figure}

When the nuclei are sufficiently small (e.g., smaller than what our system can resolve, $\sim$10$\mu$m), bubbles tend to grow with a circular projection, particularly when well-separated from other bubbles (figs.~\ref{fgr:reel_decompressionT}C, D, F, H). If the nuclei are relatively large, bubbles grow with a certain anisotropy that resembles the contour of the nucleus before approaching circularity (figs.~\ref{fgr:bubble_morphology}A-B). The root cause is that our bubbles grow with a pinned contact line as PVB adhesion with the confining glass is strong.

The morphology of the bubbles is additionally affected by their neighbors. As displayed in fig.~\ref{fgr:bubble_morphology}C, bubbles which initially grew with a circular projection eventually develop flat faces towards their neighbors, as if \textit{repelling} one another, whenever they become relatively close. Similarly, the overall bubble morphology can become highly irregular if the original nuclei are anisotropic and near one another, i.e., circularity is by no means guaranteed in our system (fig.~\ref{fgr:bubble_morphology}D). The anisotropy developed by neighboring bubbles that are sufficiently close is probably related to the polymer elasticity\textemdash however marginal at our temperatures\textemdash which prevents the bubbles from coalescing (fig.~\ref{fgr:bubble_morphology}E-F).

\section{Conclusion}

We have studied the nucleation and growth of bubbles post decompression at elevated temperatures in a model system of glass/polyvinyl butyral/glass. 

Two gaseous species are relevant in our experiments: air and water vapor.  From a thermodynamical point of view at saturation, the former promotes bubble dissolution, while the latter induces growth. For air, we have provided a brief\textemdash albeit much-needed\textemdash characterization of mass transport to help fill a pressing gap in the available literature. These results on the transport of different gases in PVB offer promising perspectives for the development of future gas separation membranes and functional films, areas in which the use of PVB is gaining in importance.

Our experimental results also show that bubble growth is directly related to the imposed temperature before decompression, i.e., the higher the temperature, the greater the likelihood of nucleating and growing a bubble. Counter-intuitively, however, the number of nuclei is negatively correlated with the imposed temperature. This may be related to the interaction between the PVB polymer and the confining glass slides. When heated, the polymer chains relax and can even reptate relative to one another, thus allowing the polymer to flow and better conform to the confining glass slides. The better flow capabilities at elevated temperatures probably improves air dissolution, thereby diminishing the number of nucleation sites available to grow a bubble. 

The nucleation of bubbles seems to always be of the heterogenous kind, i.e., bubbles grow at a polymer/glass interface. This is not to say that homogenous (bulk) nucleation is not possible. However, our accumulated body of experimental work highlights that the former is more likely than the latter. Our post-decompression bubbles grow with a morphology linked to the initial gas nuclei. When the latter is sufficiently small and well-separated from other neighboring bubbles, the bubble takes on a circular shape. If the nuclei are rather large, the bubble grows with a contour shape that initially mimics the original nuclei shape because the contact line of the bubble is always pinned. Even in the case of circular bubbles, anisotropy can ensue if bubbles become close to one another.  

Elegantly, the nucleation mechanism can be subdivided in a similar fashion to champagne’s into natural and artificial nucleation. Natural nucleation consists of bubbles nucleating and growing at a trapped dust speckle or fiber. Artifical nucleation involves bubbles forming at an imperfection in the glass, such as a pit or crenel. We have provided evidence for both. Bubbles growing from natural nucleation exhibit a constant \textit{global} growth rate, whereas those growing from artificial nucleation experience a growth rate linked to their \textit{local} conditions, in particular, the size of the defect itself. 

\begin{acknowledgments}
The authors thank Nathalie Rohaut, Guillaume Dupeux for bubbling discussions, Daniele Costantini for lending his expertise with UV laser glass patterning, and Jerome Giraud for his help in designing and developing the experimental set-up.
\end{acknowledgments}

\appendix

\section{Nitrogen sorption}\label{app:gravimetry}

\begin{figure}
\includegraphics[width=0.48\textwidth]{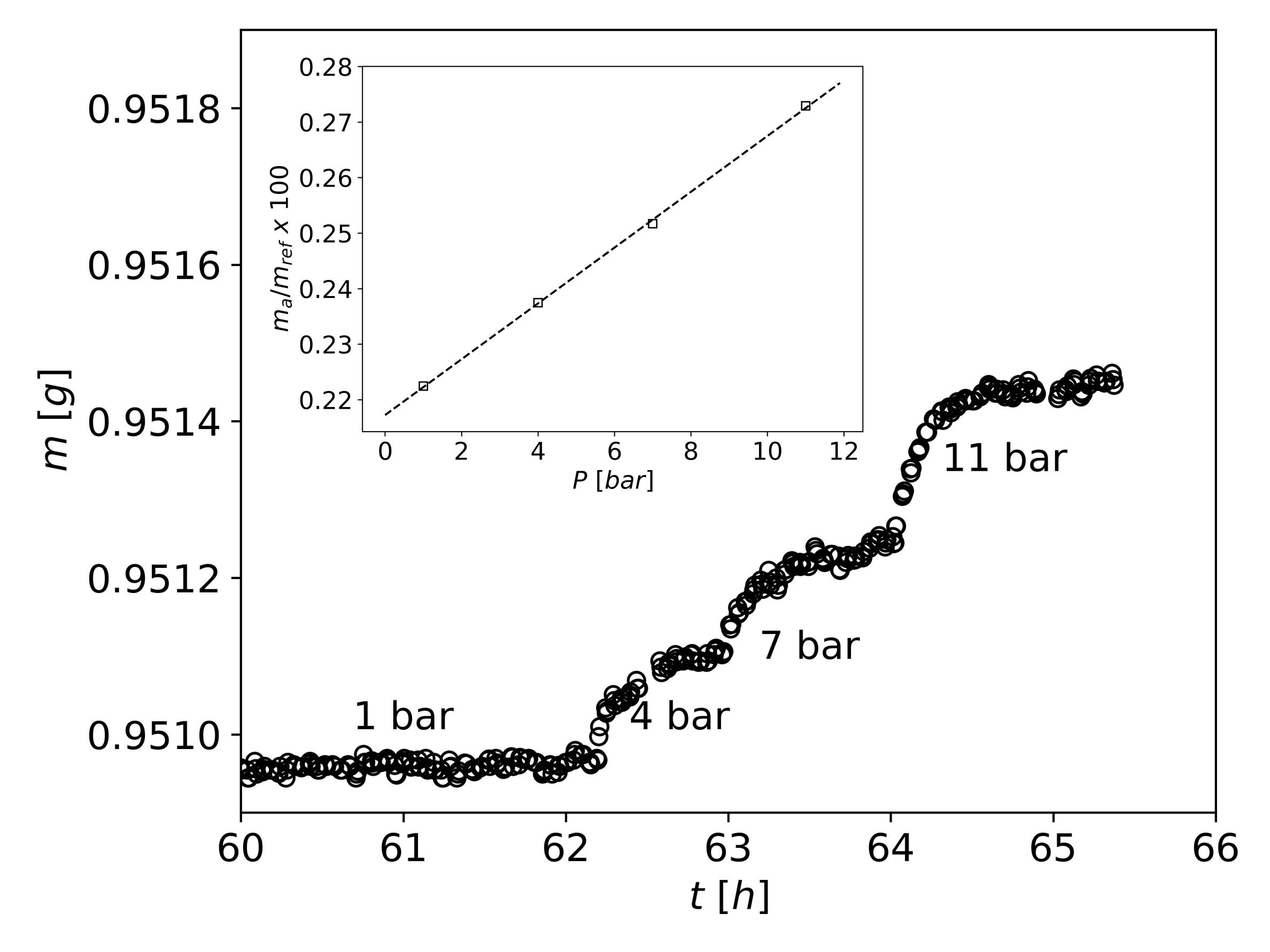}
\centering
\caption{\scalebox{1.2}{$\circ$} Transient PVB  mass during a sorption experiment for $N_2$ (20$^{\circ}$C) as a function of the applied pressure (absolute). The inset presents the sorption isotherm ({\scriptsize{$\square$}}), i.e., the steady uptake of mass $m_a$ (with buoyancy correction) with respect to the reference mass $m_{ref}$ (degassed PVB mass). \textcolor{black}{\rule[.5ex]{.5em}{.5pt}\,\rule[.5ex]{.5em}{.5pt}} Henry slope, i.e., the solubility constant of $N_2$.}
\label{fgr:air_sorp_20C}
\end{figure}

We employed gravimetric sorption analysis to determine N\textsubscript{2} transport constants (Rubotherm sorption scale) in collaboration with the Calnesis laboratory (Clermont-Ferrand, France). The N\textsubscript{2} sorption experiments included two steps (fig \ref{fgr:air_sorp_20C}): degassing by vacuum (not shown) and step-like inputs of pressure at a given temperature $T$. Diffusion coefficients were extracted from the transient uptake of mass by fitting the diffusion equation numerically at each pressure step, namely, $\frac{\partial c}{\partial t}=D \Delta c$, where $D$ is the diffusion coefficient and $c$ is the gas concentration. Meanwhile, the Henry solubility constant $H$ was determined from the slope of the isotherms by taking the slope between 1 and 11 bar, since we observed a non-zero intercept (inset in fig. \ref{fgr:air_sorp_20C}). Finally, to determine the activation energy of diffusion ($E_d$) and the heat of solution ($\Delta H_s$), we assumed a Arrhenius behaviour with temperature\cite{rogers1985permeation}:

\begin{eqnarray}
D&=&D_o e^{\frac{-E_d}{R_u T}}, \\
H&=&H_o e^{\frac{-\Delta H_s}{R_u T}.}
\label{eq:act_sol_constants}
\end{eqnarray}

Only low-temperature measurements yielded meaningful results. Starting at 60°C (not shown), the PVB mass decreased linearly with time under vacuum. No stabilization took place after 16hrs, which prevented us from having a reference mass. We took this as a hint that the PVB polymer loses volatiles or degrades in a continuous fashion when subjected to vacuum at elevated temperatures. 

Since our experiments involve temperatures above 60°C, we investigated thermal degradation further. We compared the mass loss between free and confined PVB (calendered and uncalendered) after heating in an oven (140°C, 60hrs). The former lost about 15-17\%wt and became yellow throughout. The latter lost approximately 2.5\%wt, with slight yellowing only around the exposed edges. Therefore, we concluded that degradation played no role in our bubble experiments. 

Table \ref{tbl:transport_comp} summarizes our findings for N\textsubscript{2} sorption in PVB at 20°C, while Table \ref{tbl:mass_transport} provides a complete breakdown of our results. In the absence of comparable data, we contrast our average transport findings at 20°C to those available for glassy PVB\cite{haraya1992permeation} (80\%wt butyral, 25°C, $T_g=51$°C), and simpler, rubbery homopolymers such as polyisoprene or polybutadiene (Table \ref{tbl:transport_comp}). 

When comparing the PVB blends, we find the solubility constants to be of comparable magnitude, but a large disparity was observed for the average diffusion coefficient. The latter finding is hardly surprising since the glassy state highly constrains the polymer chains, thus hindering the diffusion of molecules. Surprisingly, despite our PVB blend including three monomer units and a plasticizer molecule, the order of magnitude of the solubility constant ($\sim10^{-5}$ kg kg\textsuperscript{-1} bar\textsuperscript{-1}), diffusion coefficients ($\sim$10\textsuperscript{-11}m\textsuperscript{2}s\textsuperscript{-1}), and the activation energy for diffusion (20.6-41.9 kJ mol\textsuperscript{-1}, see table \ref{tbl:act_diff}), were all in the range observed for N\textsubscript{2} in the homopolymers of polysioprene or  polybutadiene. This highlighted to us that since $N_2$ is an inert gas, it rarely interacts with the polymer matrix, thus the similarity of its transport properties across rubber-like materials with a chemistry comprised mainly by carbon, hydrogen, and oxygen. The only quantity extracted from gravimetric analysis that did not fit into this comparison framework was the sign of the heat of solution.  Indeed, our gravimetric analysis suggested that the solubility constant of N\textsubscript{2} seemingly decreased when the temperature was raised from 20 to 40°C (See Table \ref{tbl:mass_transport}, runs A and B), which yielded \textcolor{red}{$\Delta H_{s,a}$}~$<0$. 
We believe this result to be unreliable for the following reasons. 

\begin{table}
  \caption{Comparison of N$_2$ solubility constant and diffusivity between RB41 and rubbery polymers}
  \label{tbl:transport_comp}
  \begin{tabular}{cccc}
    \hline
       & $H$ [kg kg\textsuperscript{-1} bar\textsuperscript{-1}] & $D$ [m\textsuperscript{2}s\textsuperscript{-1}]& $Ed$ [kJ/mol]\\
    \hline
     RB41 &$5.55\times 10^{-5}$&4.38$\times$10\textsuperscript{-11}& 20.6-41.9   \\
     Glassy PVB\textsuperscript{\emph{a}}
&$9.7\times 10^{-5}$& 0.13$\times$10\textsuperscript{-11} & -\\
     Polyisoprene\textsuperscript{\emph{b}}& $6.89\times 10^{-5}$&7.56$\times$10\textsuperscript{-11}&31.7\\
     Polybutadiene\textsuperscript{\emph{b}}&$6.95\times 10^{-5}$&8.88$\times$10\textsuperscript{-11}&31.1\\
    \hline
  \end{tabular}
  
  \textsuperscript{\emph{a}} \citet{haraya1992permeation}
  \textsuperscript{\emph{b}}\citet{van1950influence}.
\end{table}

First, the experimental resolution of our gravimetric analysis was not sufficient to draw any quantitative conclusion. An additional sorption analysis at 20°C (run C in Table \ref{tbl:mass_transport}) revealed a measurement variability of about 27\% (runs A and C), which is larger than the perceived difference with temperature of 12\% (runs A and B). Therefore, the measured solubility constants at 20 and 40°C are unfortunately in the uncertainty range of the experiment. 

Then, it contradicts thermodynamic expectations when considering that nitrogen is incondensable at room conditions. From a theoretical point of view, if we take the solubility process as condensation at the surface followed by mixing in the bulk, then $\Delta H_s=\Delta H_{cond}+\Delta H_1$, where the former is the heat of condensation (negative in sign) and the latter is the heat of mixing (positive in sign)\cite{van1950influence, van1946permeability, klopffer2001transport, rogers1985permeation}. For incondensable nitrogen at room conditions, it holds that $\Delta H_{cond}\ll \Delta H_1$, and evidently \textcolor{red}{$\Delta H_{s,a}$}~$>0$.

Last, our bubble experiments also suggested \textcolor{red}{$\Delta H_{s,a}$}~$>0$. Indeed, when using dry PVB, we did not observe any bubble nucleation post decompression, despite extra air being dissolved in the PVB bulk from the anisotropic bubbles that initially populate our glass samples. 

\begin{table} 
  \caption{Air transport properties in PVB deduced from gravimetric analysis}
  \label{tbl:mass_transport}
  \begin{ruledtabular}
  \begin{tabular}{cccc}
      Run & P [bar] & $H$ [kg kg\textsuperscript{-1} bar\textsuperscript{-1}] & $D$ [m\textsuperscript{2}s\textsuperscript{-1}]\\
    \hline
     A-20°C & 1 & &-\\
            & 4 & &-\\
            & 7 & &$4\times 10^{-11}$\\
            & 11 & $5.02\times10^{-5}$ &$7\times 10^{-11}$\\
    \hline
     B-40°C & 1 && -\\
            & 4 & &$8\times 10^{-11}$\\
            & 7 & &$1\times 10^{-10}$\\
            & 11 & $4.39\times10^{-5}$ &$1.2\times 10^{-10}$-\\
    \hline
     C-20°C & 1 & &$2.5\times 10^{-11}$\\
            & 11 & $6.08\times10^{-5}$ &$4\times 10^{-11}$\\
  \end{tabular}
  \end{ruledtabular}
\end{table}

\begin{table} 
  \caption{Activation energy for diffusion}
  \label{tbl:act_diff}
  \begin{ruledtabular}
  \begin{tabular}{ccc}
       P [bar] &\multicolumn{2}{c}{$E_d$ [J/mol]}\\
      & Run A & Run C\\
    \hline
    7 & $3.50\times10^4$& -\\
    11& $2.06\times10^4$ &$4.2\times10^4$\\
  \end{tabular}
  \end{ruledtabular}
\end{table}

We therefore draw the following conclusions: By analogy with simpler rubbery materials and based on our bubble growth experiments, we expect the sign and value of the heat of solution for N\textsubscript{2} in PVB to be positive and in the order of 1-10 kJ mol\textsuperscript{-1}\cite{van1950influence, van1946permeability}. 

In this paper, we took a middle of the road value of 5 kJ mol\textsuperscript{-1}. A sensitivity analysis over the 1-10 kJ mol\textsuperscript{-1} range across our experimental conditions reveals that the air relative saturation is bounded by $0.51\leq$~$\color{red}{f_a}$~$\leq1.7$, i.e., air is around the saturation point in our experiments regardless of the precise value of $\color{red}{\Delta H_{s,a}}$.

\begin{table} 
  \caption{Summary of water parameters}
  \label{tbl:model}
  \begin{ruledtabular}
  \begin{tabular}{ccc}
       T\textsubscript{hold}  & 120&140\\
    \hline
    D\textsubscript{w} [$m^2/s$] & $9.28\times10^{-10}$ &$1.43\times10^{-9}$ \\
    H\textsubscript{w} [$kg \ m^{-3} Pa^{-1}$] & $9.93\times10^{-5}$ & $5.33\times10^{-5}$\\
    H\textsubscript{w} (T=25°C) [$kg \ m^{-3} Pa^{-1}$]  &\multicolumn{2}{c}{$5.96\times10^{-3}$} \\
  \end{tabular}
  \end{ruledtabular}
\end{table}

\bibliography{Champagne_to_conf_poly}% Produces the bibliography via BibTeX.

\end{document}